\documentclass[apjl,a4paper,12pt,useAMS]{emulateapj}
\usepackage{amssymb}
\usepackage{graphicx}

\begin{document}
\title{Rapidly Evolving and Luminous Transients Driven by Newly Born Neutron Stars}
\author{Yun-Wei~Yu$^{1}$, Shao-Ze~Li$^{1}$, and Zi-Gao~Dai$^{2,3}$}

\altaffiltext{1}{Institute of Astrophysics, Central China Normal
University, Wuhan 430079, China, {yuyw@mail.ccnu.edu.cn}}
\altaffiltext{2}{School of Astronomy and Space Science, Nanjing
University, Nanjing 210093, China} \altaffiltext{3}{Key Laboratory
of Modern Astronomy and Astrophysics (Nanjing University), Ministry
of Education, China, {dzg@nju.edu.cn}}

\begin{abstract}
We provide a general analysis on the properties of emitting material
of some rapidly evolving and luminous transients discovered recently
with the Pan-STARRS1 Medium Deep Survey. It is found that these
transients are probably produced by a low-mass non-relativistic
outflow that is continuously powered by a newly born, rapidly
spinning, and highly magnetized neutron star. Such a system could
originate from an accretion-induced collapse of a white dwarf or a
merger of a neutron star-neutron star binary. Therefore,
observations to these transients would be helpful for constraining
white dwarf and neutron star physics and/or for searching and
identifying gravitational wave signals from the mergers.
\end{abstract}

\section{Introduction}
Neutron stars (NSs) are usually born from core collapse of a massive
star, during which a luminous supernova can be driven by the intense
neutrino emission from the proto-NS by expelling and heating the
stellar envelope. Besides this most popular origination, there could
be other two ways leading to a NS formation. On one hand, it is
believed that, if electron captures in the core can take place more
quickly than the nuclear burning, an accreting white dwarf (WD)
could collapse into a NS by loosing pressure support rather than
lead to a disruption event (e.g. Type Ia supernova; Canal \&
Schatzman 1976; Nomoto \& Kondo 1991). On the other hand, recently
it is often suggested that a massive NS could form and survive from
a merger of a NS-NS binary (Dai et al. 2006; Fan \& Xu 2006; Metzger
et al. 2008; Bucciantini et al. 2012; Rowlinson et al. 2010, 2013),
during which a highly-beamed relativistic jet can be launched to
produce a short-duration gamma-ray burst (GRB; Nakar 2007; Berger
2014). The clues to the post-merger NSs have emerged more and more
from the observations of extended emission, X-ray flares, and
plateau afterglows of some short GRBs.

During accretion-induced collapses of WDs and NS-NS mergers, a
low-mass non-relativistic outflow can be ejected nearly
isotropically due to the dynamical centrifugal force and the wind
blown from a remnant NS$+$disk system. In such an outflow, some
radioactive heavy elements are expected to be synthesized. Then the
decay of the elements would generate a faint optical-infrared
transient with a duration of several days and a luminosity of
$\lesssim10^{41}\rm erg~s^{-1}$ (Darbha et al. 2010; Li \&
Paczy{\'n}ski 1998; Kulkarni 2005; Rosswog 2005; Metzger et al.
2010; Roberts et al. 2011). In the NS-NS merger case, such
transients are now known as kilonovae or macronovae, which have
become one of the recent highlights of time domain astronomy since
from the observation of the infrared bump after GRB 130603B (Tanvir
et al. 2013; Berger et al. 2013).

However, as implied by the X-ray and optical afterglows of GRB
130603B, the GRB jet and thus the isotropic outflow could both be
primarily powered by the spin-down of a newly born NS rather than
radioactivities (Fan et al. 2013), although its infrared luminosity
is still too ambiguous to achieve a definite conclusion. In any
case, this observation encouraged us optimistically to consider that
the luminosity of the transients during WD collapses and NS-NS
mergers could be enhanced significantly by a spinning-down
post-collapse/merger NS to be comparable to or even exceed those of
ordinary supernova, but still with an obviously short duration due
to the low mass of the outflow (Yu et al. 2013; Metzger \& Piro
2014). Reasonably, a new class of rapidly evolving and luminous
transients were recently reported by Drout et al. (2014) with the
Pan-STARRS1 Medium Deep Survey (PS1-MDS), which could therefore be
the first smoking-gun observational signatures for the NS-powered
transient events during WD collapses and/or NS-NS mergers.

\section{The PS1-MDS transients}
The PS1-MDS transients were specifically identified by following
criteria: the transient must rise by $>$1.5 mag in the 9 days
immediately before a peak and decline by $>$1.5 mag in $\sim$25 days
after the peak. Such rapid evolution of the transients makes them
easily evade from previous supernova surveys until the PS1-MDS which
is featured in its rapid cadence and multiple band coverage to a
significant depth ($\sim$24 mag). For 10 of the new-discovered
PS1-MDS transients, the spectra of their underlying hosts were
obtained so that the cosmological redshifts of the transients can be
determined to range from $z = 0.074$ (PS1-10ah) to $z = 0.646$
(PS1-11bbq). Meanwhile, the evolution of the spectral energy
distributions of the transients selves was also achieved by
interpolating their light curves in five broadband filters ($g_{\rm
P1}r_{\rm P1}i_{\rm P1}z_{\rm P1}y_{\rm P1}$). Then, the effective
black-body temperatures of the transients (except for PS1-12bb) were
revealed to evolve around $20,000$ K, which indicates an ultraviolet
emission, near the peak emission time to 7000 K at later times. The
pseudo-bolometric luminosities of the transients were
correspondingly calculated to have a peak value with the order of
$L_{\rm p}\sim 10^{43}\rm erg~s^{-1}$. Consequently, the radius of
emission region at the peak time can be constrained to be $R_{\rm
p}=\left({L_{\rm p}/ 4\pi \sigma T_{\rm
p}^4}\right)^{1/2}\sim3.0\times10^{14}{\rm cm~}L_{\rm
p,43}^{1/2}T_{\rm p,4.3}^{-2}$, where $\sigma$ is the
Stefan-Boltzmann constant and, hereafter, the conventional notation
$Q_{x}=Q/10^x$ is adopted in the cgs units.

The above radius, luminosity, and temperature of the PS1-MDS
transients indicate that they are probably associated with
stellar-size energetic ``explosions". According to the selection
criteria of Drout et al. (2014), the peak times $t_{\rm p}$ of the
transient light curves can be taken to be from a few to $\sim10$
days after the explosions. Such short peak times make the transients
distinct from the overwhelming majority of ordinary supernovae,
which usually reach their peaks after several weeks or a few months
from the explosions. More precisely and specifically, the PS1-MDS
transients with an intermediate luminosity $<10^{43}\rm erg~s^{-1}$
(i.e. PS1-10ah, PS1-10bjp, PS1-12bb, PS1-12brf, PS1- 13dwm, and
PS1-13ess) still exhibit some similarities with the
previously-reported rare supernova events such as SN 2002bj
(Poznanski et al. 2010), SN 2010X (Kasliwal et al. 2010), and SN
2005ek (Drout et al. 2013). However, the completely new
luminosity-timescale region opened by the high-luminosity PS1-MDS
transients (i.e. PS1- 11qr, PS1-11bbq, PS1-12bv, and PS1-13duy),
which has never been probed before, would lead us to consider a
completely new branch of stellar explosions.

Regardless of the specific nature of the progenitors of the PS1-MDS
transients, a general order-of-magnitude analysis can be made with
respect to a cluster of material of mass $M$ and internal energy
$E_{\rm int}$ expanding due to its own internal pressure. On one
hand, the peak of the material emission generally appears at the
time within which photons inside the material can transfer from the
core to the outmost surface. By considering of the random walk of
the photons, the peak emission time can be expressed by $t_{\rm
p}\sim\left({R_{\rm p}/ \lambda}\right)^2{(\lambda/ c)}={3\kappa
M_{\rm }/ 4\pi R_{\rm p}c}$, where $\lambda=(\kappa\rho)^{-1}$ is
the average free path, $\kappa$ the opacity, and $\rho$ the density.
Then the mass and the kinetic energy of the material can be
estimated by
\begin{eqnarray}
M_{\rm }&\sim&{4\pi R_{\rm p}ct_{\rm p}\over 3\kappa}=0.04~{\rm
M_{\odot}}~\kappa_{-0.3}^{-1}L_{\rm p,43}^{1/2}T_{\rm
p,4.3}^{-2}t_{\rm p,6}^{}, \label{mass}
\end{eqnarray}
and
\begin{eqnarray}
E_{\rm k,p}&=&{1\over2}M_{\rm }v_{\rm p}^2\sim3.3\times10^{48}{\rm
erg}~\kappa_{-0.3}^{-1}L_{\rm p,43}^{3/2}T_{\rm p,4.3}^{-6}t_{\rm
p,6}^{-1},
\end{eqnarray}
respectively, where the expanding velocity is given by
\begin{eqnarray}
v_{\rm p}\sim{R_{\rm p}\over t_{\rm p}} =3.0\times10^8{\rm
cm~s^{-1}}~L_{\rm p,43}^{1/2}T_{\rm p,4.3}^{-2}t_{\rm p,6}^{-1}
\label{velocity}.
\end{eqnarray}
On the other hand, the emission luminosity can be approximatively
connected with the internal energy by $L\sim 4\pi RE_{\rm int}c/(3
\kappa M)$ by following the diffusion equation $f=({c}/{3 \kappa
\rho})({\partial u_{\rm int}}/{\partial r})$, where $f=L/(4\pi R^2)$
is the emission flux, $u_{\rm int}=E_{\rm int}/V$ the internal
energy density, and $V={4\over3}\pi R^3$ the volume. Then we get
\begin{eqnarray}
E_{\rm int,p}&\sim&{3\kappa ML_{\rm p}\over 4\pi R_{\rm p}c}=L_{\rm
p}t_{\rm p}=1.0\times10^{49}{\rm erg~}L_{\rm p,43}^{}t_{\rm p,6}^{}.
\end{eqnarray}
Together with the equation $4\pi R^2p/M=dv/dt\sim v/t$ describing
the adiabatic acceleration of the material, the expanding velocity
at the peak time can also be calculated by
\begin{eqnarray}
{v_{\rm p}}&\sim&{4\pi R_{\rm p}^2p_{\rm p}t_{\rm p}\over M_{\rm
}}=4.5\times10^8{\rm cm~s^{-1}}~\kappa_{-0.3}^{}T_{\rm
p,4.3}^{4}t_{\rm p,6}^{},\label{velocity2}
\end{eqnarray}
where the pressure $p={1\over3}u_{\rm int}$ is adopted. By combining
Equations (\ref{velocity}) and (\ref{velocity2}), we can obtain a
rough constraint on the opacity as $\kappa\sim0.7~{\rm
cm^2~g^{-1}}~L_{\rm p,43}^{1/2}T_{\rm p,4.3}^{-6}t_{\rm p,6}^{-2}$,
which indicates an opacity primarily contributed by electron
scattering. The above constraint also implies that the basic
properties (e.g. mass, expanding velocity, and internal energy) of
adiabatically expanding material can be certainly determined with
the peak values of its bolometric light curve.

\begin{figure*}
\centering\resizebox{0.7\hsize}{!}{\includegraphics{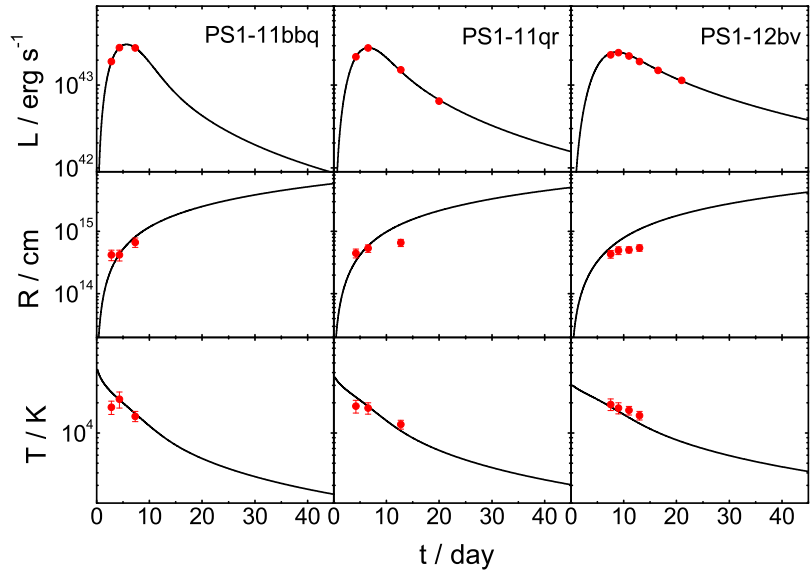}}\caption{Comparison
between the model-predicted (solid lines) and the observational
pseudo-bolometric luminosity, radius, and temperature of the
high-luminosity PS1-MDS transients (solid circles). The source
PS1-13duy is excluded because only two data points were provided.
}\label{fitting}
\end{figure*}
\section{Energy engine }
The mass of the emitting material, which is dramatically smaller
than those of ordinary supernova ejecta, suggests that the PS1-MDS
transients probably involve an explosion of a system governed by an
extremely-stripped star or, more probably, stellar-size compact
objects. The material ejected during the explosion could be heated
by the energy engine ``impulsively" or by some other energy sources
after the explosion.

First of all, as a conventional consideration, the emitting material
could be impulsively heated by a shock and therefore the emission
luminosity is determined by the shock breakout radius. However,
owing to a limit on the radius of the largest red supergiant
(Levesque et al. 2009), only the PS1-MDS transients with an
intermediate luminosity $<10^{43}\rm erg~s^{-1}$ can possibly be
attributed to a shock breakout from a normal star. For the
high-luminosity ones (i.e. PS1- 11qr, PS1-11bbq, PS1-12bv, and
PS1-13duy), Drout et al. (2014) suggested that these transients
could be associated with a shock breakout within a dense stellar
wind. In this case, however, an unexpected high mass-loss rate and
an ad hoc mass-loss history of the progenitor need to be invoked.
Then, Kashiyama \& Quataert (2015) proposed that the impulsive
heating could be accomplished in a very hot super-Eddington
fall-back accretion disk surrounding a newly born black hole, where
a low-mass outflow can be driven by the disk radiation pressure.
Such a model can work if the collapsing progenitor star is
elaborately designed to eject a just-right mass of material so that
a successful supernova can be avoided but an appropriate fall-back
accretion disk can still be maintained.

\begin{table}
\centering
\renewcommand{\arraystretch}{2.0}\caption{Parameters for fittings to the light curves}
\begin{tabular}{llll}
 \hline
 \hline
 Events & $M/\rm 10^{-2}M_{\odot}$ & $\xi L_{\rm sd,i}/10^{44}\rm erg~s^{-1}$ & $t_{\rm sd}/\rm day$
\\\hline
 PS1 - 11bbq  & 2.0 & 4.5 & 2.0
\\
 PS1 - 11qr  & 1.5 & 1.5 & 5.0
 \\
PS1 - 12bv  & 1.5 & 0.65 & 14\\
\hline \hline
\end{tabular}
\end{table}
In contrast, a post-explosion energy supply could be a more natural
and attractive choice, which can be provided by radioactive decay of
heavy elements and/or a long-lasting energy engine. In the
radioactivity scenario, the mass of heavy elements represented by
$^{56}\rm Ni$ is required to be $M_{\rm Ni}\sim m_{\rm Ni}{E_{\rm
int,p}/( q_{\rm Ni}+q_{\rm Co})}\sim0.05~{\rm M_{\odot}}L_{\rm
p,43}t_{\rm p,6}$, where $m_{\rm Ni}$ is the mass of a $^{56}$Ni
atom and $q_{\rm Ni}=$1.78 MeV and $q_{\rm Co}=$3.505 MeV are the
energy release for the decay of per $^{56}$Ni and per subsequent
$^{56}$Co, respectively. As shown, for a sufficiently high
luminosity, the required nickel mass could unrealistically exceed
the total mass of the emitting material. Even for a relatively low
luminosity, an unacceptable high nickel fraction is also predicted,
because the nickel mass is only slightly lower than the total mass.
Therefore, the radioactivity must not be the primary energy source
for the PS1-MDS transients and alternatively a long-lasting energy
engine becomes necessary and crucial. The long activity of the
engine further requires that the post-explosion compact object
should be a NS rather than a black hole, as previously suggested for
GRBs (Dai \& Lu 1998a,b; Zhang \& M\'{e}sz\'{a}ros 2001) and
superluminous supernovae (Kasen \& Bildsten 2010). The energy that a
NS can provide to power the explosion outflow mainly comes from its
spin-down with a luminosity $L_{\rm sd}(t)=L_{\rm
sd,i}\left(1+{t}/{t_{\rm sd}}\right)^{-2}$. Then by taking
appropriate values of the parameters $M$, $L_{\rm sd,i}$ and $t_{\rm
sd}$ as listed in Table 1, we employ a semi-analytical light curve
model (see Appendix for details) to account for the
pseudo-bolometric light curves of the high-luminosity PS1-MDS
transients (i.e., PS1-11qr, PS1-11bbq, and PS1-12bv) as well as the
evolution of radius and temperature of the transients, as presented
in Figure \ref{fitting}. In our fittings, a typical constant opacity
$\kappa=0.5\rm cm^2g^{-1}$ is artificially adopted to reduce the
parameter freedom and an uncertain parameter $\xi$ is introduced to
represent the energy injection efficiency.

Being represented by the braking mechanism due to magnetic dipole
radiation, the spin-down luminosity and timescale of a NS can be
expressed by $L_{\rm sd,i}=10^{7}{\rm ~erg~s^{-1}}~B_{}^{2}P_{\rm
i}^{-4}$ and $t_{\rm sd}=2\times10^{39}{\rm s}~B_{}^{-2}P_{\rm
i}^{2}$, where $B$ and $P_{\rm i}$ are the magnetic field strength
and initial spin period, respectively. Then, from the fitting
parameters, the values of $B$ and $P_{\rm i}$ can be derived as
functions of the parameter $\xi$, as presented in Table 2. It is
shown that the involved NSs should be highly magnetized and spin
rapidly with a millisecond period, in spite of the uncertainty of
the energy injection efficiency. In short, the PS1-MDS transients
(at least the high-luminosity ones) are probably produced by a
low-mass outflow that is continuously powered by a millisecond
magnetar. Specifically, such a newly born NS could originate from an
accretion-induced collapse of a WD or a merger of a NS-NS binary.
During these processes, a strong magnetic field can be naturally
generated and amplified via flux freezing and dynamo action (Duncan
\& Thompson 1992; Cheng \& Yu 2014). On the contrary, for an
extremely-stripped star, it is nearly impossible to produce a
millisecond magnetar through its collapse.

\begin{table}
\centering
\renewcommand{\arraystretch}{2.0}\caption{Neutron star parameters}
\begin{tabular}{llll}
 \hline
 \hline
 Events & $P_{\rm i}/\rm ms$ & $B/10^{14}\rm G$
\\\hline
 PS1 - 11bbq & 16$\xi^{1/2}$ &17$\xi^{1/2}$
\\
 PS1 - 11qr & 18$\xi^{1/2}$  & 12$\xi^{1/2}$
\\
 PS1 - 12bv & 16$\xi^{1/2}$ & 6.5$\xi^{1/2}$
\\
\hline \hline
\end{tabular}
\end{table}

\section{Discussions}
In view of the high similarity in the post-explosion configurations
between the WD collapses and NS-NS mergers, it is not easy to
clearly discriminate these two models by the present observations of
the PS1-MDS transients.

Firstly, the derived spin periods of $\sim16-18$ ms for $\xi\sim 1$
seems only prohibited by WD collapses but conflict with NS-NS
mergers, because the high initial angular momentum of the binary
would give a limiting spin period of $P_{\rm i}\sim 1$ ms to the
post-merger NSs. However, in fact, it is widely believed that such a
near-Keplerian spinning NS would lose its rotational energy
immediately by radiating gravitational waves (Dall'Osso et al.
2015), amplifying magnetic fields (Cheng \& Yu 2014), and producing
short GRBs and subsequent extended emission (Metzger et al. 2008;
Bucciantini et al. 2012). As a result, the initial period for
magnetic dipole radiation would become much longer than $\sim1$ ms,
e.g., $P_{\rm i}=5$ ms (e.g. Cheng \& Yu 2014). For such a reference
value, we can constrain the energy injection efficiency to be
$\xi\sim10$\%. This might be caused by that (1) the majority of the
spin-down luminosity is concentrated into the GRB direction to
contribute a plateau afterglow emission (Yu et al. 2010), which
however probably deviates from the light of sight and, perhaps, (2)
the spin-down luminosity on the light of sight could also be partly
released through X-ray emission by the NS wind itself (Dai 2004;
Metzger \& Piro 2014). Therefore, such wind X-ray emission
associated with the UV-optical transients, if exists, could provide
a testable signal for our model.

Secondly, the NS-NS merger model is seemingly challenged by other
two issues. (1) As pointed out by Drout et al. (2014), the explosion
site offsets of all transients with respect to their host galaxies
are statistically smaller than those expected for NS-NS mergers. (2)
As calculated by Kasen et al. (2013), the opacity of the
neutron-rich outflow could be on the order of magnitude of
$10-100\rm cm^2~g^{-1}$ due to the synthesization of lanthanides,
which is much higher than the opacity inferred here. However, on one
hand, if we individually pick up the offsets of the high-luminosity
transients, we will find that most of them are actually on the high
side of the offset distribution. On the other hand, the opacity of
the outflow actually can be reduced by two effects: (1) the
lanthanide synthesization in the disk wind component of the outflow
is blocked due to the neutrino irradiation from the post-merger NS
(Metzger \& Fern{\'a}ndez 2014) and (2) the lanthanides in the
dynamical component of the outflow could be ionized by the X-ray
emission from the NS wind (Metzger \& Piro 2014).

Finally, a potential discrimination between the two models may arise
from the explanation of the event rate of PS1-MDS transients, i.e.,
roughly, $\sim3200\rm Gpc^{-3} yr^{-1}$ corresponding to the four
high-luminosity transients. As a comparison, the rate of Type Ia
supernovae was found to $(3.01\pm 0.062) \times 10^4 \rm Gpc^{-3}
yr^{-1}$ from the Lick Observatory Supernova Search (Li et al. 2011)
and the rate of NS-NS mergers is also widely considered to be on the
order of $\sim1000\rm Gpc^{-3} yr^{-1}$ (Kalogera et al. 2004).
Therefore, optimistically, the event rate of PS1-MDS transients
could in principle be accounted for in both the WD collapse and
NS-NS merger models, as long as a remarkable fraction of these
processes can lead to a NS formation. The rationality of such
fractions could be tested by some further model implications, e.g.,
the cosmic fraction of neutron-rich elements, the detectability of
associated gravitational wave signals by Advanced LIGO and Virgo
(Abadie et al. 2010; Nissanke et al. 2013), and the detectability of
subsequent radio transients arising from the interaction between the
outflow and circum medium (Metzger et al. 2015), etc. For this
purpose, some detailed calculations and simulations need to be
implemented with the constraints derived in this Letter.

\section{Conclusion}
The luminosity-timescale region opened by the PS1-MDS transients,
which although could be partly overlapped by the shock breakout
emission of some supernovae, leads us to get a glimpse of a
completely new branch of stellar-size explosion phenomena. The mass
and energy requirements on the emitting material for the transients
robustly indicate a post-explosion system consisting of a newly born
millisecond magnetar and a low-mass non-relativistic isotropic
outflow. Such a system could originate from accretion-induced
collapses of WDs or mergers of NS-NS binaries. Therefore, future
observations to these transients would play a vital role in
constraining WD and NS physics and/or in searching and identifying
gravitational wave radiation.

\acknowledgements We thank M. R. Drout for sharing their data.
Y.W.Y. thanks Professor K. S. Cheng and The University of Hong Kong
for hospitality while this work was initiated. This work is
supported by the National Basic Research Program of China (973
Program, grant 2014CB845800), the National Natural Science
Foundation of China (grant No. 11473008), and the Program for New
Century Excellent Talents in University (grant No. NCET-13-0822).

\appendix
\section{The light curve model}
\begin{figure}
\centering\resizebox{0.7\hsize}{!}{\includegraphics{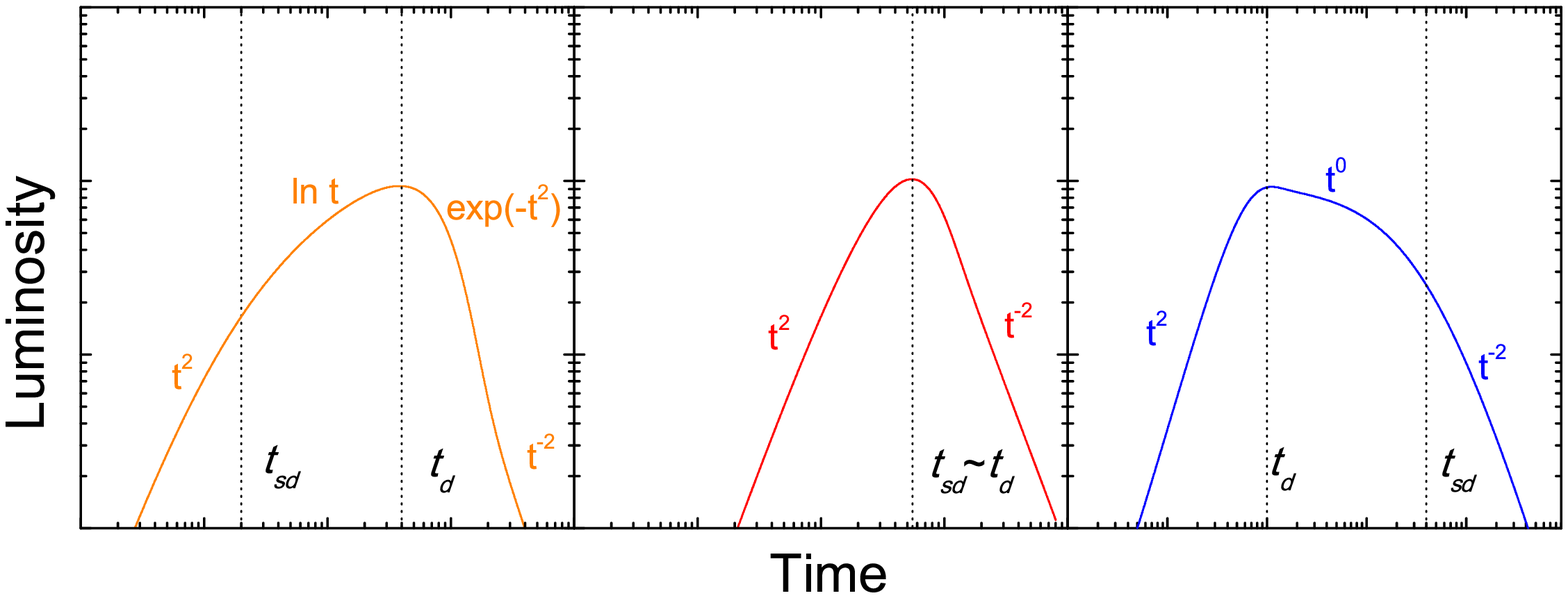}}\caption{Three
typical types of light curves, where the approximative temporal
behaviors are labeled for different evolution phases.}\label{LC1}
\end{figure}

Following Arnett (1980) and Kasen \& Bildsten (2010), a
semi-analytical light curve model for NS-powered mergernovae was
provided by Yu et al. (2013), where relativistic effects were taken
into account. Here, in view of the non-relativistic velocity
inferred from the PS1-MDS transients, we alternatively adopt a
simplified non-relativistic version of the model, which was
previously described in detail by Kasen \& Bildsten (2010) for
superluminous supernovae. In this case, the dynamical evolution of
an outflow can be determined by
\begin{eqnarray}
{dv_{\rm }\over dt}={4\pi R^2p\over M_{\rm }}
\end{eqnarray}
and
\begin{eqnarray}
{dR\over dt}=v_{\rm },
\end{eqnarray}
where the pressure $p={1\over3}(u_{\rm int}-u_{\rm e})$ with $u_{\rm
int}$ being the internal energy density and $u_{\rm e}$ being the
energy density releasing by emission. With an energy supply rate
$\xi L_{\rm sd}$, the evolution of the outflow internal energy is
determined by the energy conservation law as
\begin{eqnarray}
\frac{ d E_{\rm int}}{d t} =  \xi L_{\rm sd} - L_{\rm }- p \frac{ d
V}{d t},\label{Eint1}
\end{eqnarray}
where $L$ is the emission luminosity and the work $-pdV=-4\pi R^2pv$
represents the energy loss due to the expansion of the outflow. The
emission luminosity can further be related to the internal energy by
$L_{\rm }={E_{\rm int}c/ (\tau R)}$ for $t\leq t_\tau$ and $L_{\rm
}={E_{\rm int}c/ R}$ for $t>t_\tau$, where $\tau=3\kappa M/4\pi R^2$
is the optical depth and the time $t_{\tau}$ represents the time for
the optical thick-thin transition. Then, by setting the parameters
$\kappa$, $M$, $\xi L_{\rm sd,i}$, and $t_{\rm sd}$, a light curve
can be solved from the above equations. Three typical types of light
curves are presented in Figure \ref{LC1}.

For an analytical understanding of the light curve profiles, we can
approximate the spin-down power of the NS by a broken-power-law
function as
\begin{eqnarray}
L_{\rm sd} &=&\xi L_{\rm d,i}, {\rm
~~~~~~~~~~~~~~~~~for~} t<t_{\rm sd},\\
&=&\xi L_{\rm d,i}(t/t_{\rm sd})^{-2}, {\rm ~~~for~~~} t>t_{\rm sd}.
\end{eqnarray}
Meanwhile, by introducing the photon diffusion time of $t_{\rm
d}=(3\kappa M/4\pi v c)^{1/2}$ with a constant velocity, Equation
(\ref{Eint1}) can be rewritten to (Kasen \& Bildsten 2010)
\begin{eqnarray}
{dE_{\rm int}t\over dt}={\dot{E}_{\rm supl}t_{\rm d}^2 -E_{\rm
int}t\over t_{\rm d}^2}t\label{Eint3}
\end{eqnarray}
for $t<t_{\tau}$. The shape of light curves mainly depends on the
relation between the timescales $t_{\rm d}$ and $t_{\rm sd}$. Let us
consider the case of $t_{\rm sd}<t_{\rm d}$ as an example. Firstly,
for $t<t_{\rm sd}$, the energy supply is constant and the solution
of Eq. (\ref{Eint3}) can be derived to
\begin{eqnarray}
L_{\rm }={E_{\rm int}t\over t_{\rm d}^2}=\xi L_{\rm
d,i}\left(1-e^{-t^2/2t_{\rm d}^2}\right)\approx\xi L_{\rm
d,i}\left({t^2\over 2t_{\rm d}^2}\right).
\end{eqnarray}
Secondly, for $t_{\rm sd}<t<t_{\rm d}$, by omitting the minor
important radiation effect and simplifying Eq. (\ref{Eint3}) to
${d(E_{\rm int}t)/ dt}\approx{\xi L_{\rm d,i}t_{\rm sd}^2 / t}$,
we can get
\begin{eqnarray}
L_{\rm }&\approx& \xi L_{\rm d,i}\left(1-e^{-{t_{\rm sd}^2/ 2t_{\rm
d}^2}}\right)+{\xi L_{\rm d,i}t_{\rm sd}^2\over t_{\rm
d}^2}\ln{t_{\rm }\over t_{\rm sd}}\nonumber\\       
&\approx&{\xi L_{\rm d,i}t_{\rm sd}^2\over t_{\rm
d}^2}\left({1\over2}+\ln{t_{\rm }\over t_{\rm sd}}\right).
\end{eqnarray}
Thirdly, for $t>t_{\rm d}$, the radiation term becomes dominative in
Eq. (\ref{Eint3}), i.e., ${d(E_{\rm int}t)/ dt}\approx -{E_{\rm
int}t^{2}/ t_{\rm d}^2}$,
which gives
\begin{eqnarray}
L_{\rm }&\approx& \left[\xi L_{\rm d,i}\left(1-e^{-{t_{\rm sd}^2/
2t_{\rm d}^2}}\right)+{\xi L_{\rm d,i}t_{\rm sd}^2\over t_{\rm
d}^2}\ln{t_{\rm d}\over t_{\rm
sd}}
\right]e^{-{(t^2-t_{\rm d}^2)\over 2t_{\rm d}^2}} \nonumber\\
&\approx&{\xi L_{\rm d,i}t_{\rm sd}^2\over t_{\rm
d}^2}\left({1\over2}+\ln{t_{\rm d}\over t_{\rm
sd}}\right)e^{-{(t^2-t_{\rm d}^2)\over 2t_{\rm d}^2}}.
\end{eqnarray}
Finally, as the emission luminosity deceases to approach the energy
supply luminosity, the succeeding evolution of the emission
luminosity should track the evolution of the energy supply (i.e.,
$L_{\rm }\propto t^{-2}$), because after $t_{\rm d}$ the freshly
deposited internal energy can escape from the outflow nearly
immediately. The above treatment can be easily extended to the other
two cases of $t_{\rm sd}\sim t_{\rm d}$ and $t_{\rm sd}> t_{\rm d}$.
Then a completed solution to Eq. (\ref{Eint3}) can be summarized as
follows:

\textit{{\rm Case I: For} $t_{\rm sd}<t_{\rm d}$}
\begin{equation}
L_{\rm }\propto\left\{
\begin{array}{ll}
t^{2},&{~\rm for~}t<t_{\rm sd},\\
\ln t,&{~\rm for~}t_{\rm sd}<t<t_{\rm d},\\
e^{-t^2},&{~\rm for~}t_{\rm d}<t<t_{\rm c},\\
t^{-2},&{~\rm for~}t>t_{\rm c},
\end{array}\right.
\end{equation}

\textit{{\rm Case II: For} $t_{\rm sd}\sim t_{\rm d}$}
\begin{equation}
L_{\rm }\propto\left\{
\begin{array}{ll}
t^{2},&{~\rm for~}t<t_{\rm sd},\\
t^{-2},&{~\rm for~}t>t_{\rm sd},
\end{array}\right.
\end{equation}

\textit{{\rm Case III: For} $t_{\rm sd}>t_{\rm d}$}
\begin{equation}
L_{\rm }\propto\left\{
\begin{array}{ll}
t^{2},&{~\rm for~}t<t_{\rm d},\\
t^{0},&{~\rm for~}t_{\rm d}<t<t_{\rm sd},\\
t^{-2},&{~\rm for~}t>t_{\rm sd},
\end{array}\right.
\end{equation}
where $t_{\rm c}$ is defined as the time at which $L_{\rm
}\rightarrow \xi L_{\rm sd}$. The above three different solutions
just correspond to the three types of light curves presented in
Figure \ref{LC1}.
In all cases, the peak
luminosity of the emission appears at the diffusion time $t_{\rm
d}$. The three high-luminosity PS1-MDS transients are all fitted by
the model with $t_{\rm sd}\sim t_{\rm d}$ in the text.

\end{document}